# Top-down fabrication of atomic patterns in twisted bilayer graphene

*Ondrej Dyck[1], Sinchul Yeom[3], Andrew R. Lupini[1], Jacob L. Swett[2], Dale Hensley[1], Mina Yoon[3], Stephen Jesse[1]*

[1] *Center for Nanophase Materials Sciences, Oak Ridge National Laboratory, Oak Ridge, TN*

[2] *Biodesign Institute, Arizona State University, Tempe, AZ 87287*

[3] *Materials Science and Technology Division, Oak Ridge National Laboratory, Oak Ridge, TN*

**Abstract**

Atomic-scale engineering typically involves bottom-up approaches, leveraging parameters such as temperature, partial pressures, and chemical affinity to promote spontaneous arrangement of atoms. These parameters are applied globally, resulting in atomic scale features scattered probabilistically throughout the material. In a top-down approach, different regions of the material are exposed to different parameters resulting in structural changes varying on the scale of the resolution. In this work, we combine the application of global and local parameters in an aberration corrected scanning transmission electron microscope (STEM) to demonstrate atomic scale precision patterning of atoms in twisted bilayer graphene. The focused electron beam is used to define attachment points for foreign atoms through the controlled ejection of carbon atoms from the graphene lattice. The sample environment is staged with nearby source materials, such that the sample temperature can induce migration of the source atoms across the sample surface. Under these conditions, the electron-beam (top-down) enables carbon atoms in the graphene to be replaced spontaneously by diffusing adatoms (bottom-up). Using image-based feedback-control, arbitrary patterns of atoms and atom clusters are attached to the twisted bilayer graphene with limited human interaction. The role of substrate temperature on adatom and vacancy diffusion is explored by first-principles simulations.

**Introduction**

The goal of any fabrication process is to structure matter in a predefined way. The physical limit to such structuring of matter is governed by the chemical bonding of individual atoms. The field of chemical engineering grapples with the challenge of rearranging molecular structures through spontaneous, albeit understood and intended, chemical reactions. Likewise, the field of material growth leverages surface chemistry to direct atoms to specific bonding sites. These approaches are typically thought of as forms of synthesis, implying the addition of various constituents to spontaneously produce a new structure. The governing chemistry ensures that structural changes occur at the atomic scale. This spontaneous restructuring is considered a bottom-up approach.

In contrast, top-down methods seek to alter materials through the application of an outside influence. Technically speaking, this is also governed by interactions at the atomic level, but from a practical perspective, the physical footprint of the outside influence defines the physical dimensions on which the technique can operate. Photolithography reaches a limit with the wavelength of light.[1] Electron-beam (e-beam) lithography leverages the smaller wavelength of electrons to achieve higher resolution. The limit of e-beam lithography is determined no longer by wavelength, but primarily by various proximity effects between the e-beam and the sample.[2] The emission of secondary electrons from the sample can also degrade the resolution. Both lithography techniques rely on masking, exposing, etching/growth, etc. to define the features. Direct material deposition strategies also exist. Electron (or ion) beam induced deposition, EBID (or IBID), relies on the interactions with an organometallic precursor gas which the e-beam dissociates.[3] EBID is typically performed in a scanning electron microscope (SEM) where a gas injection system is used to flow the precursor gas across the workpiece. Here too, emission of secondary and backscattered electrons from the substrate can limit the resolution.

An enhancement in resolution may be achieved by performing EBID in a scanning transmission electron microscope (STEM) where the reduction in the e-beam size as well as a reduction in substrate thickness both act to improve resolution. Van Dorp et al.[4–8] performed landmark demonstrations toward moving EBID from an SEM into a STEM, where aberration correction and thin samples enabled finer-scale beam profiles and less substrate scattering. They employed a single layer of graphene as their substrate, which is atomically thin and has a very low SE yield[9–12] and were ultimately able to show the addition of material down to the molecular level. It was emphasized by the authors that when approaching this level of precision, the sample cleanliness was pivotal. What is lacking, however, and is neglected in descriptions of EBID generally, is an atomic scale conceptualization of what occurs as the molecule attaches to the sample surface. These details have been understandably absent since EBID is typically performed on scales much larger than single atoms and molecules, where the material under consideration can be treated as continuous. In contrast, synthesis methods are primarily concerned with the atomistic details.

Typically, EBID is performed using an organometallic precursor gas that introduces unwanted organic species onto the sample and deposition sites.[13] Strategies such as molecular beam epitaxy (MBE) avoid this problem by direct evaporation or sublimation of the solid source material, which leads to high purity deposition.[14] However, only limited attempts have been made to precisely direct where the source material attaches to the substrate. As top-down methods edge into atomic scale territory, questions of the atomistic details must be addressed. EBID cannot be performed with single atoms of source material since it relies on the chemical reactivity of the dissociated organic fragments to bond with the sample surface. MBE delivers purified material, and as such, there can be no reactive dissociation. To merge these paradigms there is another option, which we explore in the current work. We use the e-beam to alter the sample surface in such a way as to induce a reaction with the source atoms when they arrive on the substrate surface.

We have previously shown that dopant atoms may be inserted into the graphene lattice by creating a defect and subsequently sputtering foreign atoms into the defect.[15–19] This technique appears generalizable in that many different source elements could be inserted into the graphene; however, the distance from the source material was limited to a few nanometers. In a previous publication[20] we also broached the topic of contamination and sample cleanliness from the perspective of STEM EBID and suggested a method for preventing the ingress of unwanted hydrocarbon contaminants from elsewhere on the sample. Notably, the contaminants were found to migrate along the graphene surfaces and not through the vacuum, which enabled e-beam-deposited barriers to be sufficient to prevent further contamination. These results set the stage for obtaining and maintaining clean substrate material and demonstrating the e-beam controlled atomic substitution process.

Parallel investigations in much the same spirit are worth highlighting. The deterministic movement of Si dopants through a graphene lattice using an e-beam has been demonstrated[15,21–23] and movement of Si dopants through the walls of a single walled carbon nanotube has been shown.[24] The assembly of primitive multi-atom structures has been demonstrated.[16] These demonstrations focus on e-beam induced processes that conserve the number of atoms within the structure, but can controllably induce a structural transformation. Implantation and STEM investigations of P,[25,26] Ge,[27] In,[28] and N[29] have successfully expanded the scope of tailored functional structures that can be realized. Global e-beam irradiation and subsequent heating of the source material has also been used to attach foreign atoms to graphene and study the spin states of single atoms.[30] A more comprehensive review of the literature in this field was also published.[31] These demonstrations illustrate the growing scientific curiosity around atom-by-atom manipulation using focused e-beams and the variety of approaches that have been used to successfully insert dopant atoms in graphene for subsequent examination.

What is lacking in this context is a demonstration of how a top-down fabrication workflow can be employed that would transform these proof-of-concept results into a more robust process. What is necessary is the ability to take a given input and produce a predictable output that can be repeated at great length with reproducible results. Perhaps the closest demonstration of this kind is the movement of three-fold coordinated Si atoms using the e-beam;[15,16,21–24] however, this process is limited by either the eventual loss of the Si atom through spontaneous graphene healing or the ejection of a C atom resulting in the much less mobile four-fold coordinated Si atom. Moreover, it is unclear whether this process generalizes to other elements and other structures.

In this work we show the proof-of-principle that a combined top-down and bottom-up atomic-scale e-beam manufacturing process is possible. The sample state is arranged such that top-down, focused e-beam exposure leads to the spontaneous, bottom-up attachment of dopant atoms or small clusters to the workpiece, twisted bilayer graphene (TBG). This process is automated so that patterning can proceed with limited human interaction and is reproducible across several days and in different instruments.

Specifically, we demonstrate the patterning of dopant atoms and small atom clusters in TBG at a variety of temperatures ranging from 1050 to 1150 °C. This was accomplished using the focused e-beam in a STEM to define the attachment locations in the TBG. The e-beam generates defects in the TBG that provide attachment sites for the foreign atoms. The sample temperature plays a primary role in controlling the foreign atom supply rate. Defect diffusion also plays a non-trivial role, particularly for the precision of positioning dopants.

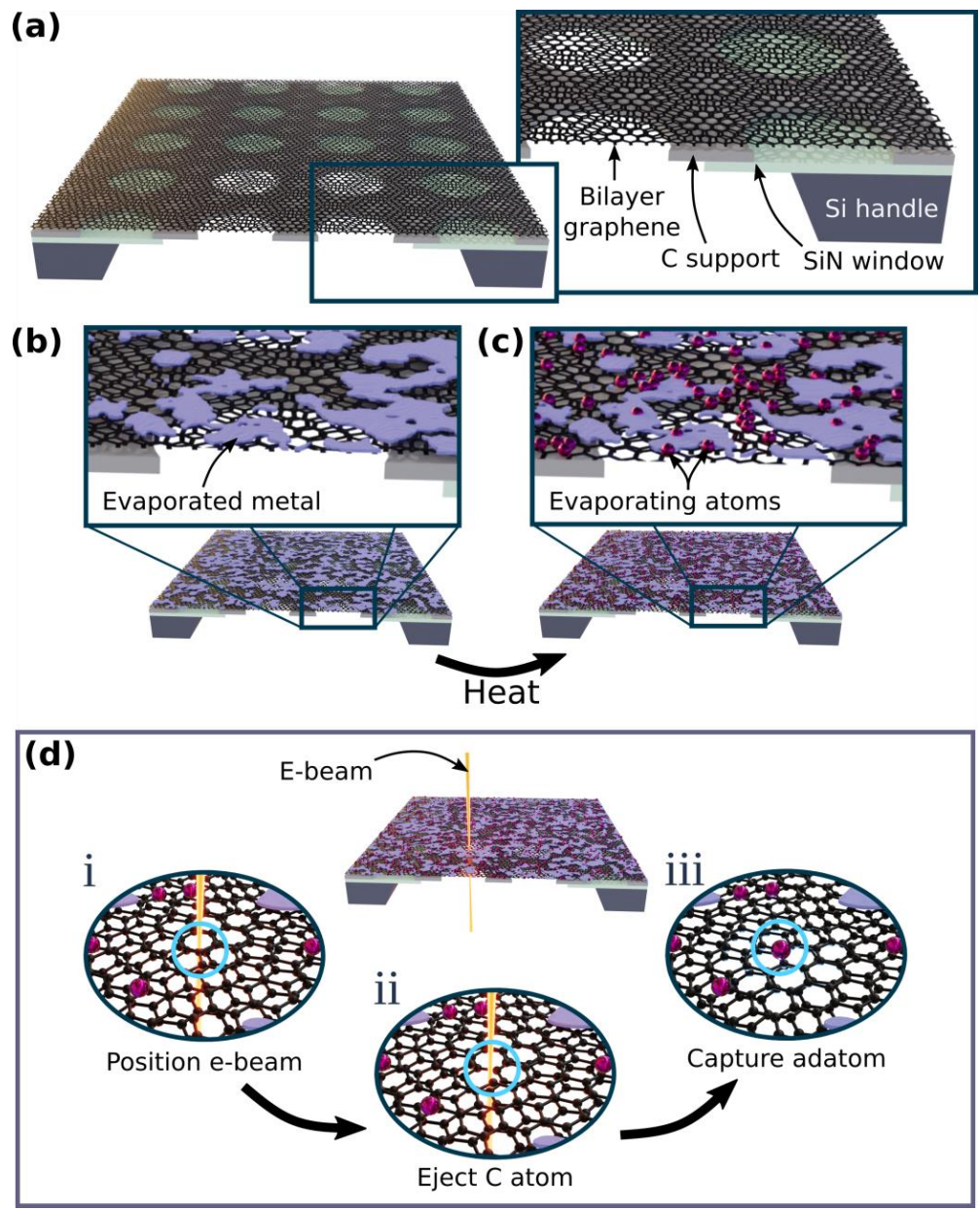

**Figure 1 Conceptual diagram showing sample preparation and dopant insertion strategy.** (a) Diagram of TBG suspended across holes in a Protochips[TM] heater chip. (b) Sample diagram after evaporating metal onto the surface. (c) Diagram of the sample during operation of the heater chip, illustrating *in situ* evaporation and diffusion of metal atoms across TBG surface. (d) Conceptual workflow for attaching diffusing adatoms to TBG involving i) positioning the e-beam at the location of interest, ii) ejection of a carbon atom, and iii) spontaneous capture of an adatom at the defect site.

**Experimental Concept**

Figure 1(a)-(c) shows a cut-away conceptual drawing depicting the sample geometry. Figure 1(a) shows the TBG on the supporting Protochips™ heater chip (not to scale). Figure 1(b) shows a conceptual drawing of the sample after source material has been evaporated onto the TBG surface. In the STEM, we heat the sample to various temperatures, which enables the evaporation and migration of single atoms of the source material across the TBG surface. Figure 1(c) shows a conceptual drawing of this process.

Figure 1(d) summarizes the e-beam dopant insertion process. The 100 keV e-beam is focused onto a target location. After some time, a carbon atom is ejected from the lattice due to direct knock-on energy transfer from the beam electrons. This leaves a vacancy, which forms an attachment site for diffusing adatoms. Since these adatoms are being continuously supplied through thermal evaporation of the source material, attachment and incorporation into the defect site proceeds without further intervention.

**Sample Overview**

After loading the sample into the microscope, the heater chip was ramped to 1200 °C at a ramp rate of 1000 °C/ms to clean/remove residual hydrocarbon contaminants from the sample surface. Figure 2(a) shows an overview high angle annular dark field (HAADF) STEM image with the major features labeled. The carbon support substrate can be seen on the lower left side of the image. The rest of the field of view consists of nanoparticles on TBG suspended over the hole in the carbon support substrate. Figure 2(b) and 2(c) show electron energy loss spectra (EELS) of the Cr and Cu $L_{2,3}$ edges acquired on the parent nanoparticles, respectively.

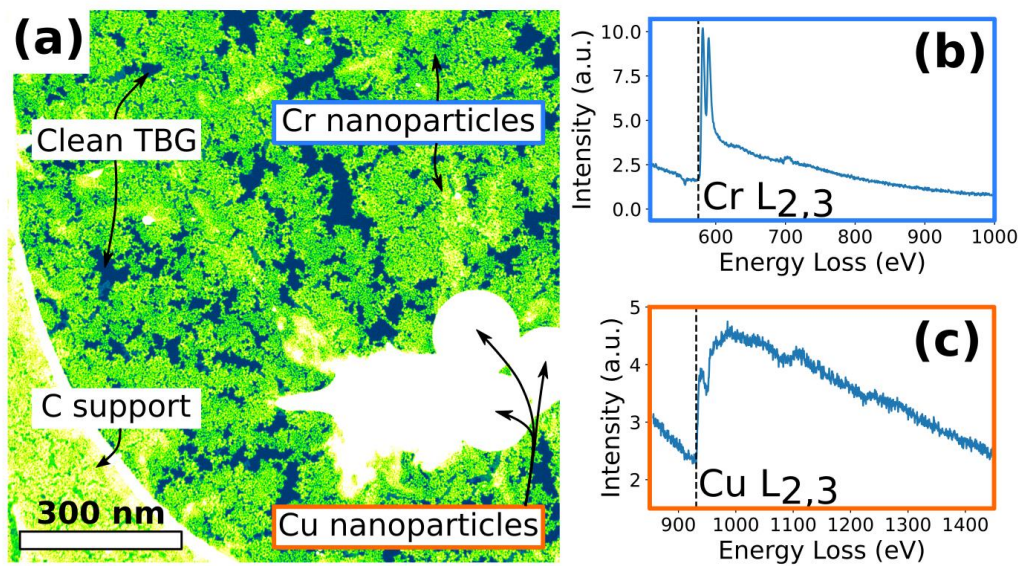

**Figure 2 Sample overview.** (a) HAADF-STEM overview image of the sample. Various major features are labeled. (b) and (c) EELS spectra used to establish the elemental identity of the nanoparticles.

**Beam Control and Patterning**

A custom-built feedback and beam control platform (described in previous publications[32–35]) was used to perform automated dopant patterning. Briefly, this platform interfaces with the microscope scan coils to control the e-beam position and concurrently samples the detector output so that it can reproduce standard image acquisition. However, with customized beam control, arbitrary patterns can be scanned and parameters such as the pixel dwell time can be varied dynamically. In this application, a small (~ 0.5 nm) spiral scan was used to define the "mill location," the location where the beam will spend an extended period to induce defect formation. This spiral pattern enables concurrent imaging, providing an approximately real-time view of the local sample state. The mean intensity of the spiral image is used as the feedback control mechanism with thresholds set both above and below the mean, which are variables set by the user during operation. The high threshold is triggered when a strongly scattering dopant attaches to the TBG and increases the observed intensity. The low threshold is triggered when a hole forms at the mill location. After either threshold is triggered, the algorithm proceeds to the next mill location. Figure 3 shows a summary of the two types of patterns used and an example of both a dopant insertion as well as the creation of a hole at the mill location.

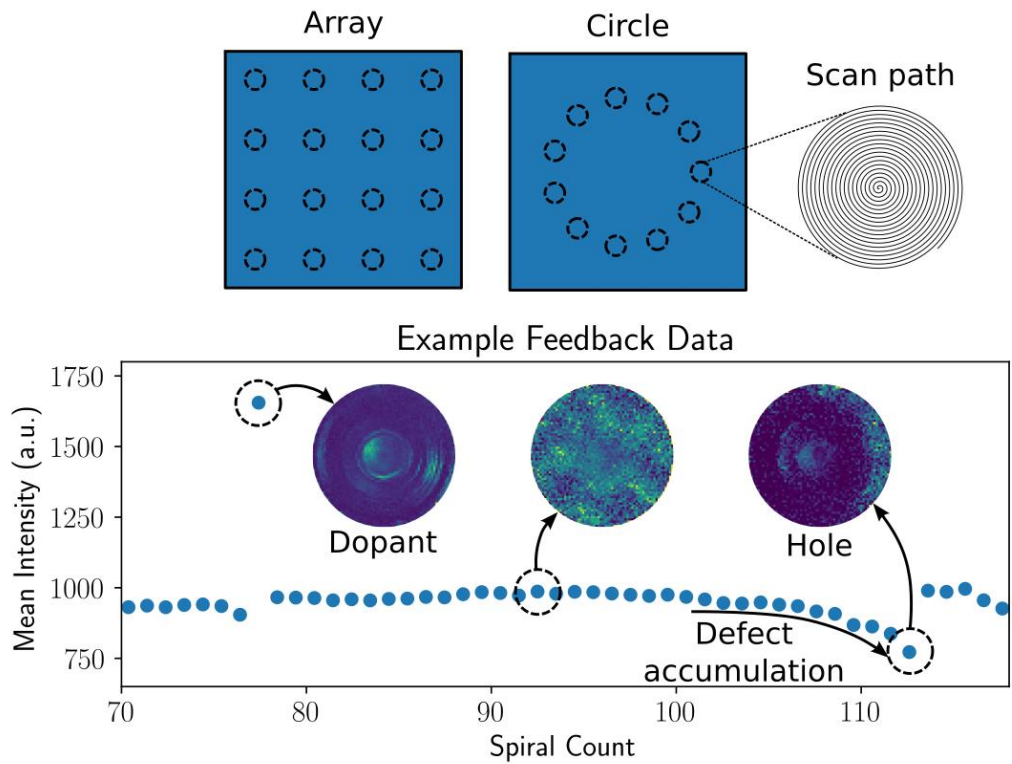

**Figure 3 Example patterns and feedback data.** Two scan patterns, arrays and circles, were used, as illustrated schematically at the top. At each position a spiral scan was performed, which provides an updating image of the local structure. An example of the progression through time is shown on the bottom. The mean intensity value of the spiral scan is plotted sequentially. 'Spiral Count' refers to the index label associated with each image. Once an outcome is obtained, either a dopant or hole, as defined by an upper and lower threshold, the algorithm moves to the next position in the pattern.

Since the 100 kV accelerating voltage continued to damage the formed structures after they were patterned, the accelerating voltage was lowered to 60 kV for subsequent imaging. Figure 4 shows representative examples of a patterned circle (a) and array (b). When attempting to establish the elemental identity of the inserted atoms using EELS, it was found that both Cr and Cu $L_{2,3}$ edges were present. The spectrum shown in Figure 4(c) was acquired while scanning over a small, nonrepresentative cluster selected specifically because it contained both Cr and Cu atoms. Discrimination between the atomic species can be obtained by comparing the HAADF-STEM intensity as illustrated in Figure 4(d). Example intensity line profiles are shown inset for easier comparison. The colored circles correspond to the different elements. While both Cu and Cr (and Si) atoms could be found, the vast majority of the dopant atoms were Cu.

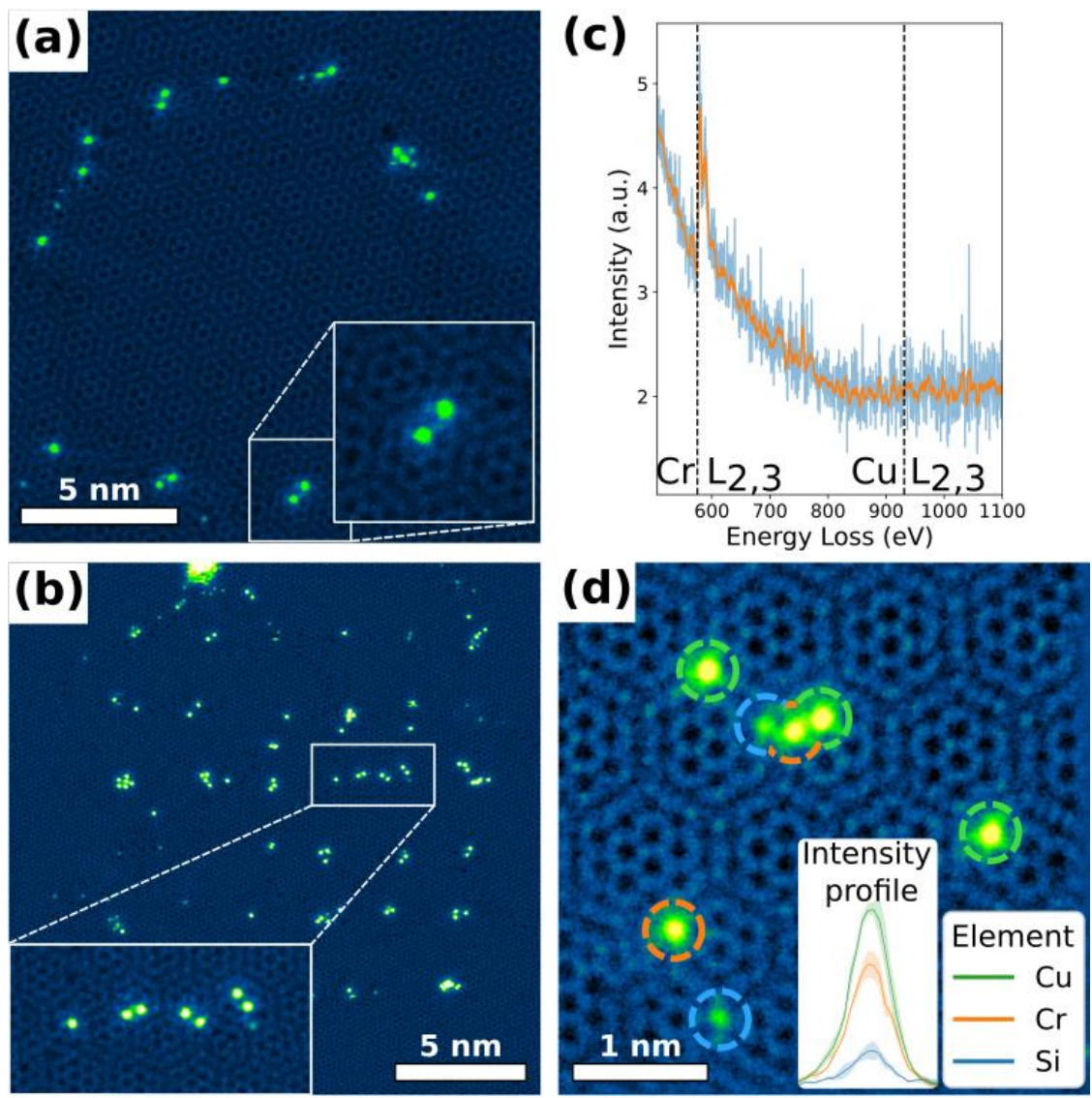

**Figure 4 Representative examples of dopant patterning and identification of elemental identities acquired at 60 kV.** (a) Patterned circle. (b) Patterned array. (c) EELS spectrum acquired while scanning the beam over a small cluster of dopant atoms. We observe both Cr and Cu $L_{2,3}$ edges. Blue is the raw data, orange represents a rolling average. (d) Example HAADF-STEM intensity comparison. Discrimination between elements can be performed based on image intensity. Example line profiles are shown inset.

**Effects of Temperature**

Raising the sample temperature has several consequences. First, it evaporates unwanted hydrocarbon contamination from the graphene surface,[20,36–38] which is a critical concern for fabrication at this scale.

Second, it promotes the spontaneous diffusion of adatoms from the parent nanoparticles that can then migrate across the surface and bond with the undercoordinated carbon atoms generated by e-beam ejection. Third, the increase in temperature significantly affects the vacancy diffusion rate in the graphene,[39] which will be discussed later. Here, we note that this vacancy diffusion is a likely cause of imprecision in the array creation. After a vacancy is created there is some probability for it to diffuse before capturing an adatom. This diffusion suggests that an improvement in positioning precision could be obtained by either lowering the temperature of the graphene or increasing the temperature of the source material to increase the supply of adatoms (or both). Clearly, to balance these competing demands, the source material and the graphene should be separated to allow independent temperature control.

Here, we varied the sample temperature to observe the dopant insertion behavior around the melting temperature of bulk Cu, nominally 1085 °C.

Dopant insertion was performed at 1050, 1075, 1100, 1110, 1115, 1125, and 1150 °C. Figure 5 shows a summary of these results. Figure 5(a) shows a histogram capturing the distribution of milling times (all temperatures together) required to obtain an outcome, either a hole milled or a dopant inserted. We found that the Birnbaum-Saunders (or fatigue-life) distribution[40] captured the shape of our observations well, which is shown as the dotted line fit to the histogram. In addition, the Birnbaum-Saunders distribution has two constraints that align with the physics of this situation, namely that the independent variable must be positive (an outcome must occur after e-beam exposure) and that the distribution must go to zero for no exposure. The Birnbaum-Saunders distribution has been developed and used for modeling the failure rate due to the accumulation of damage, so it is perhaps unsurprising that it provides a very nice fit to our data. The inset shows a plot of the hazard function, which represents the instantaneous failure rate through time. Here, "failure rate" indicates the rate at which an event (dopant or hole) is achieved. The hazard function has a distinct peak at the beginning and converges asymptotically to a constant value determined by the shape, $\alpha$, and scale, $\beta$, parameters.[41] This value is expressed as $1/(2\alpha^2\beta)$ and evaluates to 0.05 s$^{-1}$ for the fit shown.

The data can be separated according to temperature and a similar fitting performed for each temperature, as shown in the supplemental information (SI). We find a slight correlation with temperature, suggesting faster outcomes at higher temperatures, which is consistent with an increased supply of source atoms. Likewise, one may also ask whether the fit changes appreciably with the outcome, dopant, or hole. This fit is also shown in the SI, where we did not find a significant difference. These observations are consistent with the understanding that the accumulation of damage at the irradiated area is responsible for either outcome.

In Figure 5(b) we show a bar chart of the outcomes observed. By the nature of the experiment, waiting until an outcome occurs ensures that either a dopant or a hole will eventually appear. The bar chart then captures the fraction of both dopants and holes at each temperature. The error bars represent the standard deviation. At 1100 °C, roughly half of the outcomes were holes and the other half were dopants. Above this temperature only dopants were obtained. Notably, the onset of the transition from obtaining some holes to obtaining only dopants occurs just above the melting temperature for bulk Cu (1085 °C). This transition is consistent with the earlier observation that most of the dopants were Cu and not Cr, even though the Cr nanoparticles were closer to the region of interest. The melting point for bulk Cr exceeds 3000 °C. Clearly, the spontaneous ejection of atoms from the source material plays a critical role in facilitating a favorable sample environment for patterning of this kind.

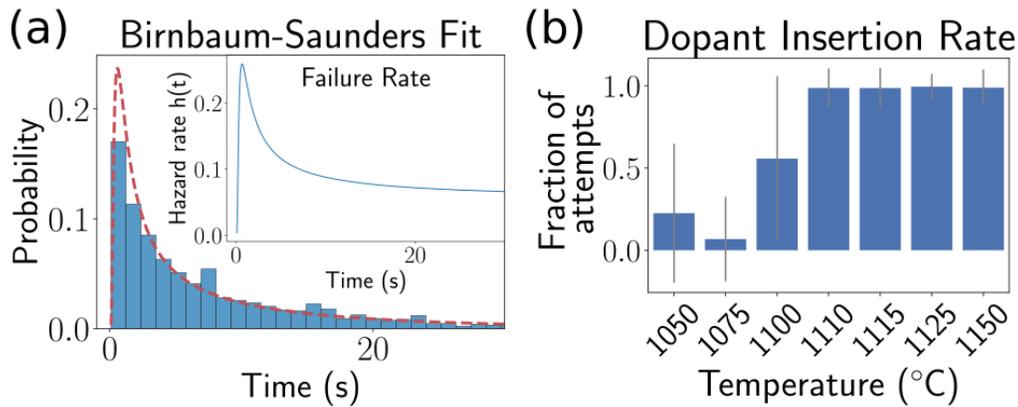

**Figure 5 The role of temperature.** (a) Density histogram showing the distribution of milling times required to obtain an outcome, either a dopant or hole, for all temperatures. The shape of the distribution closely matches the Birnbaum-Saunders (fatigue-life) distribution; optimal fit is shown by the dotted overlay. (b) Bar chart showing fraction of outcomes that resulted in a dopant (rather than a hole) as a function of temperature. Error bars represent the standard deviation.

**Vacancy Formation, Diffusion, and Metal Insertion – Experiments and Theoretical Calculations**

Experimentally, we cannot easily separate the role of vacancy diffusion within the TBG from the rate of supply of source atoms without independent control of the temperature of the source material and the TBG. Nevertheless, we can examine whether vacancy diffusion plays a role in the rate of obtaining an outcome. Evidence for the diffusion of defects away from the mill location can be seen in the array shown in Figure 4(b), where defect chains are formed between mill locations in the TBG and dopant atoms are found in unintended locations. In a previous publication the milling rate of single layer graphene as a function of temperature was explored and a significant role for vacancy diffusion was found.[39] Specifically, the higher the vacancy diffusion rate, the higher the electron dose required to mill a hole. This effect was pronounced at high temperatures and at 1000 °C the average dose needed to mill a hole exceeded $1.8 \times 10^{11}$ e/nm$^2$. We note that the ability to mill holes in TBG at temperatures around 1000 °C suggests that the vacancy diffusion rate is suppressed by the presence of a second layer. The mean dose required to obtain an outcome here, averaged across all temperatures, was $6.4 \times 10^9$ e/nm$^2$. Counterintuitively, bilayer graphene requires significantly less dose to mill at ~1000 °C than monolayer graphene. Because this effect is governed by vacancy migration, these dynamics can be examined theoretically providing a grounding for the experimental results.

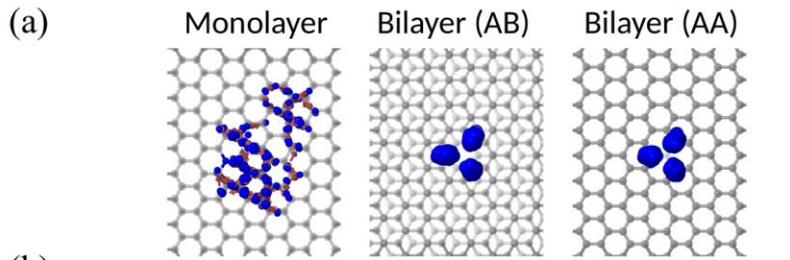

| Binding Energies (eV) | Gr | AB | AA |
|---|---|---|---|
| Top | -0.5427 | | |
| Hollow | -0.3902 | -0.4865 | -0.5291 |
| Bridge | -0.5366 | -0.6196 | -0.6501 |
| Monovacancy | -4.2111 | -4.0355 | -4.0660 |
| Vacancy on Hollow | | -4.0236 | |
| Divacancy | -5.6688 | -5.8537 | -5.8488 |

|  | Gr | AB | AA |
|---|---|---|---|
| Top | ✓ | | |
| Hollow | ✓ | ✓ | ✓ |
| Bridge | ✓ | ✓ | ✓ |
| Monovacancy | ✓ | ✓ | ✓ |
| Vacancy on Hollow | | ✓ | |
| Divacancy | ✓ | ✓ | ✓ |

**Figure 6** (a) Molecular dynamics simulations at 2000 K show rapid diffusion of defects on monolayer graphene, while the diffusion process is suppressed due to interlayer interactions in bilayer graphene with both AA and AB stacking. (b) Table of binding energies for Cu atoms in various graphene structures are summarized in the bottom table.

Molecular dynamics (MD) simulations were performed to model the vacancy diffusion dynamics in monolayer and bilayer graphene. Figure 6 shows the diffusion dynamics of a monovacancy in monolayer graphene and bilayer graphene with AA and AB stacking at 2000 K. For the monolayer example, the vacancy location is shown through time as a blue ball with brown arrows indicating the trajectory. The diffusion coefficient was $3.197 \times 10^7$ nm$^2$/s (for details see ref.[39]). For the bilayer graphene examples, the blue balls represent the undercoordinated C atoms that define the edge of the vacancy. The stability of the edge atoms results in a substantial suppression of vacancy diffusion and only local vibrations are observed. This tendency persists up to ~2500 K. Interlayer van der Waals interactions between the graphene planes stabilize the vacancy in the bilayer examples, agreeing well with the experimental observations. At each temperature the simulation time was 10 ns.

We also performed first-principles density functional theory (DFT) calculations to evaluate the energetics of graphene capturing Cu adatoms using FHI-aims software. The binding energy of Cu on monolayer pristine graphene is ~0.54 eV (top, hollow, and bridge sites). This increases to 0.62 eV and 0.65 eV for AB and AA stacking graphene, respectively, i.e., the Cu adatom binding energy increases with the bilayer stacking, as summarized in the tables in

Figure 6(b). The decrease in binding energy observed for the bilayer systems is induced by a strong local distortion in the neighborhood of the Cu dopant atom, indicating that the mechanism for incorporating Cu adatoms in the graphene system is governed by the kinetics.

To determine whether the effects of vacancy diffusion are still present, we examined the positions of the mill locations within the arrays of holes. The arrays were milled in a raster pattern beginning at the top left and ending at the bottom right. As the array is created, different positions have different numbers of neighbor locations that received prior irradiation by the e-beam. Therefore, if vacancy diffusion contributes in a meaningful manner to the processes under examination, we should be able to detect a difference in time to an event for the different locations. For simplicity, we consider first and second nearest neighbors (NN) and assume that the effects of more distant mill locations are negligible. The categories described next are summarized in Figure 7(a). The first mill location has no NNs, forming the first category. The top row of mill locations, except for the first, have one NN on the left. The start of each row after completion of the first row has one NN (above) and one second NN (SNN) (above and to the right), forming category three. The end of each row, after completion of the first row, has two NNs (above; left) as well as one SNN (above and to the left), forming the fourth category. Finally, locations within the central portion of the array have two NNs (above; left) as well as two SNNs (above and to the right; above and to the left). The category labels 1-5 have been chosen such that the number of NNs and SNNs increases with the label index. By dividing the data into these five categories, we can fit a Birnbaum-Saunders distribution to each category and examine whether there is a notable change in the shape of the distribution captured by the shape parameter, $\alpha$ (for a discussion on how this parameter effects the distribution see the SI). Figure 7(b) shows a plot of the shape parameter for each category. A linear fit is provided as a guide for the eye. The dots are color coded according to category and sized proportional to the number of data points represented. A clear downward trend is observed. In Figure 7(c) the corresponding hazard functions are plotted for each category and it is observed that the overall instantaneous failure rate (the rate at which we obtain an outcome) increases with the category number.

Taken together, these results suggest that vacancy diffusion from previous mill locations increases the speed with which an outcome can be achieved. However, the effect is limited to short distances (the NN distance, center-to-center, was nominally 4.3 nm).

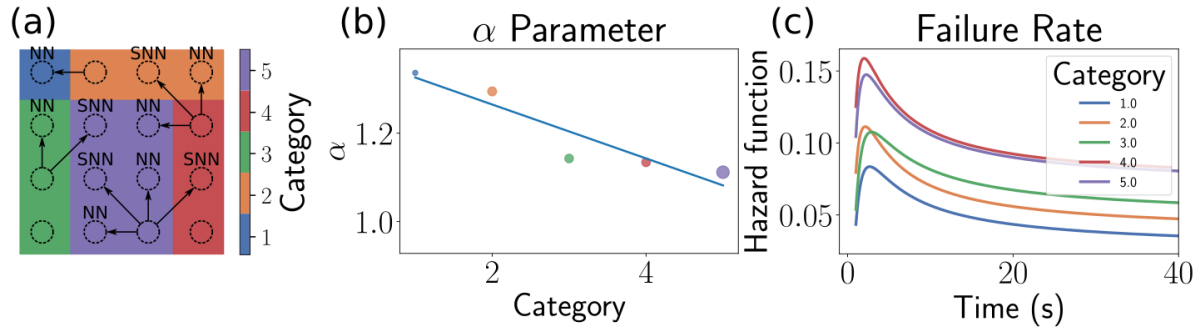

**Figure 7 Examination of nearest neighbor influence on event rate.** (a) Illustration of the different categories within an array that is created in a raster fashion from the top left to the bottom right. Category 1 has no nearest neighbors as it is the first location. Category 2 has one nearest neighbor. Category 3 has one nearest neighbor and one second nearest neighbor. Category 4 has two nearest neighbors and one second nearest neighbor. Category 5 has two nearest neighbors and two second nearest neighbors. (b) The shape parameter is plotted as a function of category. The size of the dots is scaled by the number of data points within that category. (c) Plot of the hazard function for each category. The addition of a nearest neighbor can be seen in the difference between category 1 and 2 or between category 3 and 4. In contrast, the effect of adding a second nearest neighbor can be seen in the difference between category 2 and 3 or between category 4 and 5.

**Conclusion**

In this work we demonstrated top-down, automated patterning of dopant atoms in TBG using an e-beam. The temperature of the sample functions as a key operational parameter for controlling the environment. Based on the significant change in outcome around the Cu melting temperature, we conclude that melting the source material significantly enhances the supply of source atoms.

This work represents a significant advance in atomic scale fabrication techniques using e-beams. In particular, we highlight the critical role of the global sample environment coupled with the precision of the e-beam that independently control the experimental outcome and the spatial position. In the present work, we did not attempt to align the dopant arrays with pre-existing features of the TBG, but there seems to be no fundamental obstacle. Likewise, extension to elements other than Cu seems plausible given the wide range of dopants that have previously been inserted into graphene. Our calculations revealed the reason for the strong suppression of vacancy diffusion in bilayer graphene, making it an excellent platform for e-beam patterning with dopants. One challenge may be the temperature needed for evaporation of the source material; however, this does not seem insurmountable for many elements.

**Methods**

**Sample preparation**

Low-pressure chemical vapor deposition (LP-CVD) was used to grow graphene on Cu foil.[28] The graphene was spin coated with poly(methyl methacrylate) (PMMA) to protect the graphene and serve as a mechanical stabilizer during substrate transfer. The Cu foil was dissolved in a bath of ammonium persulfate and deionized (DI) water. After rinsing in a bath of DI water, the sample was scooped onto a Protochips™ heater chip and baked on a hot plate at 150 °C for 15 minutes to promote adhesion of the graphene to the chip surface. The PMMA was subsequently removed in a bath of acetone and the chip was then dipped in isopropyl alcohol and dried in air.

Nominally 3 Å of Cr (Kurt Lesker) was evaporated onto the graphene surface using a Thermionics VE-240 e-beam evaporator operated with a deposition rate of ~0.1 Å/s. A quartz crystal microbalance in combination with a shutter was used to precisely control the amount of deposition. We hypothesize that the nanoparticles of Cu observed on the sample were from incomplete etching of the Cu substrate.

**STEM Examination**

Prior to experimentation in the STEM, the sample, cartridge, and magazine were baked in vacuum at 160 °C for 8 hrs.

A Nion UltraSTEM 100 was used to examine the sample. The accelerating voltage used for dopant insertion was 100 kV with a beam current of 120-130 pA. The accelerating voltage used for higher resolution imaging and EELS was 60 as noted in the text with a beam current of 20 pA. The nominal convergence angle was 30 mrad.

**MD Simulations**

Molecular dynamics simulations were performed to compare vacancy diffusion in single and bilayer graphene using Large-scale Atomic/Molecular Massively Parallel (LAMMPS)[14] based on Adaptive Intermolecular Reactive Empirical Bond Order - Morse (AIREBO-M)[42] empirical potential. Three different graphene cases were compared: single layer, AB stacked (layer distance 3.30 Å), and AA stacked bi-layer (layer distance 3.33 Å). Each single sheet was ~200 Å in diameter (~ 11,000 atoms), and the layer distance was determined by minimizing the system's potential energy as a function of layer distance. A vacancy was created in the middle of the top layer and 10 ns of canonical ensemble (NVT) simulations were performed at three temperatures, 1500, 2000, and 2500 K. The atomic trajectories were recorded every 25 ps. In each trajectory, carbon atoms whose number of nearest neighbors are two or four are marked as blue and their overlapped trajectories at 2000 K are shown in Figure 7 (a). The brown arrows indicate the movement of vacancies.

Fritz Haber Institute ab initio materials simulations (FHI-aims),[43–45] an all-electron code with localized numerical orbitals as the basis was used, and tight basis sets and Perdew-Burke-Ernzerhof (PBE) functional[46,47] was used. (*vdW_correction_Hirshfeld*)

We obtained a fully relaxed graphene sheet (6o carbon atoms) with CC bond length is 1.422 Å and unit cell is 12.315 x 12.798 x 40 Å$^3$ periodic box (rectangular cell), 4x4x1 k-points. Using this graphene, we created other configurations in Figure 7(b) and relaxed them. The obtained AB and AA bilayer graphene's layer distances are 3.26 Å and 3.41 Å, respectively. The Broyden-Fletcher-Goldfarb-Shanno (BFGS)[48] algorithm was used for the relaxation until the maximum force component became less than 5x10$^{-3}$ eV/Å.

**Acknowledgement**

This work was supported by the U.S. Department of Energy, Office of Science, Basic Energy Sciences, Materials Sciences and Engineering Division (O.D. A.R.L., S.J., M.Y., S.Y.), and was performed at the Center for Nanophase Materials Sciences (CNMS), a U.S. Department of Energy, Office of Science User Facility (O.D.). This research used resources of the Oak Ridge Leadership Computing Facility and the National Energy Research Scientific Computing Center, US Department of Energy Office of Science User Facilities.

**Supplemental Information**

The data shown in Figure 5(a) of the main text shows a histogram of the time required to achieve an outcome, either a dopant or hole. This dataset included all temperatures. Here, we separate the data into each temperature category and fit a Birnbaum-Saunders distribution to the resulting histogram. The results are shown in Figure 1. The shape parameter for each fit is displayed as an overlay, as well as the number of datapoints represented. In the lower right panel the shape parameters are plotted as a function of temperature. The size of the datapoint markers corresponds to the number of datapoints contained in the dataset. An unweighted linear fit is provided as a guide to the eye. A slight increasing linear trend is observed which suggests that increasing sample temperature results in faster outcomes. This is consistent with an increasing supply of Cu adatoms and the transition from creating holes to inserting dopants. However, examining the histograms, we see that only the dataset acquired at 1100 °C contained sufficient samples to produce a smooth distribution. The rest of the datasets had various levels of 'noise' roughly correlating to the number of samples in the dataset. While we expect to observe some change in the shape of the distribution of outcomes with temperature it is unclear whether the number of datapoints is sufficiently large to reliably capture these changes.

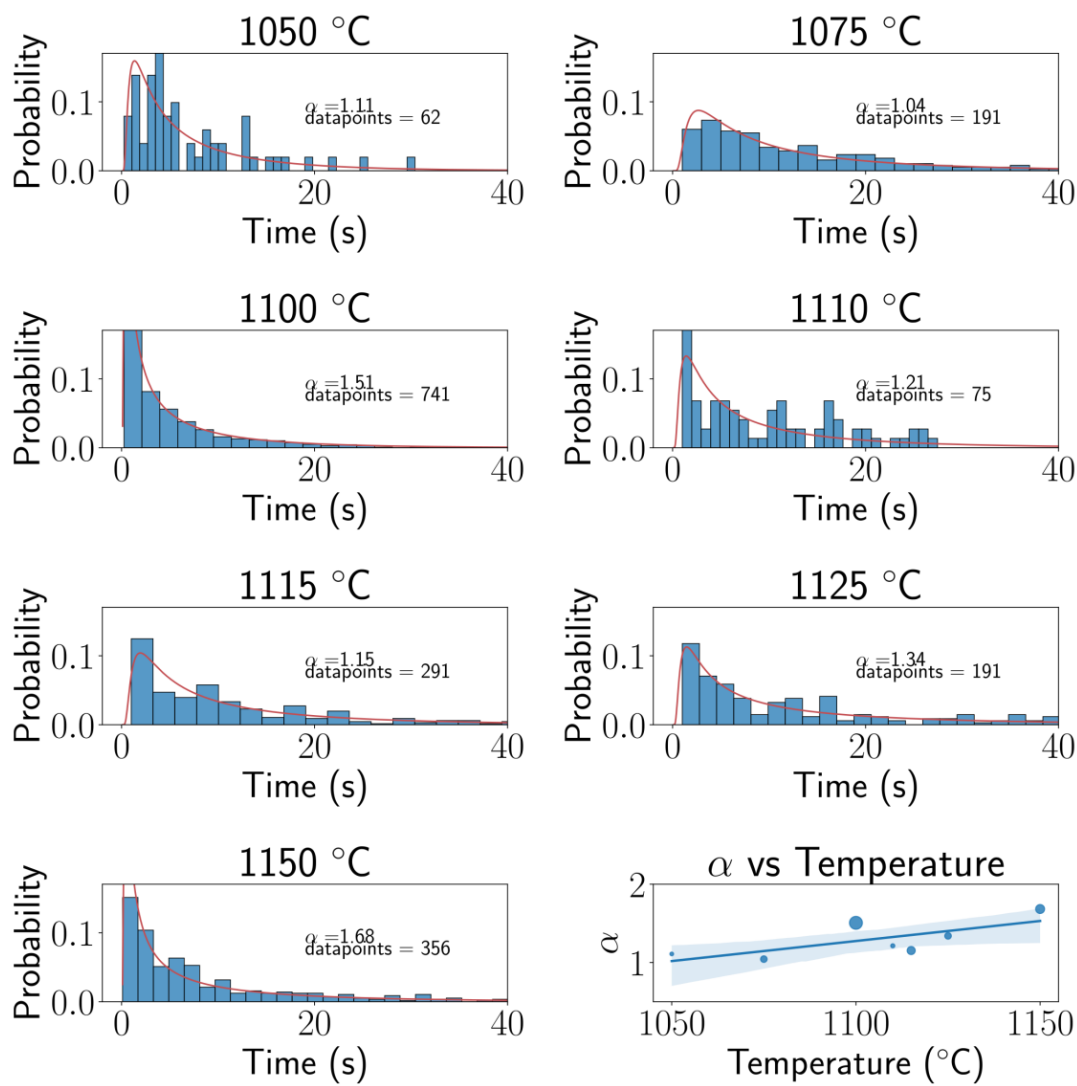

**Figure S8 Time to outcome distributions for each temperature.** A Birnbaum-Saunders distribution was fit to each histogram. The shape parameter for the fit is overlaid as well as the number of datapoints contained within each dataset. In the lower right panel, the shape parameters are plotted as a function of temperature.

Similarly, we may also separate the data to compare whether there is a difference in the shape of the distribution depending on which outcome was observed. This analysis is shown in Figure 2. We see that there is no significant difference between the two outcomes which suggests that the same physical process is driving both outcomes.

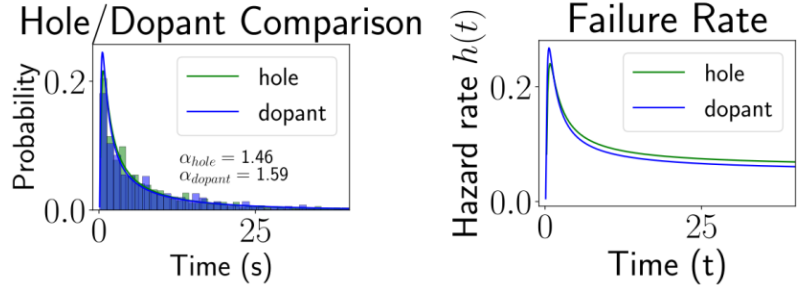

**Figure S9 Comparison between the distribution of holes and dopants.** Birnbaum-Saunders distributions were fit to each. The shape parameter is overlaid. The hazard rates are compared on the right.

The Birnbaum-Saunders (BS) probability density function is given by[1]

$$f_T(t; \alpha, \beta) = \frac{1}{2\sqrt{2\pi}\alpha\beta} \left[ \left(\frac{\beta}{t}\right)^{1/2} + \left(\frac{\beta}{t}\right)^{3/2} \right] exp\left[ -\frac{1}{2\alpha^2}\left(\frac{t}{\beta} + \frac{\beta}{t} - 2\right) \right]$$

$\alpha$ is the shape parameter and $\beta$ is the scale parameter. To illustrate how variations in these parameters affect the distribution, in Figure 3 we plot various values for $\alpha$ with $\beta = 1$ and various values of $\beta$ with $\alpha = 1$. The left panels show the as-calculated distributions and the right panels show the same distributions rescaled to a maximum of one for easier direct comparison.

## Shape parameter

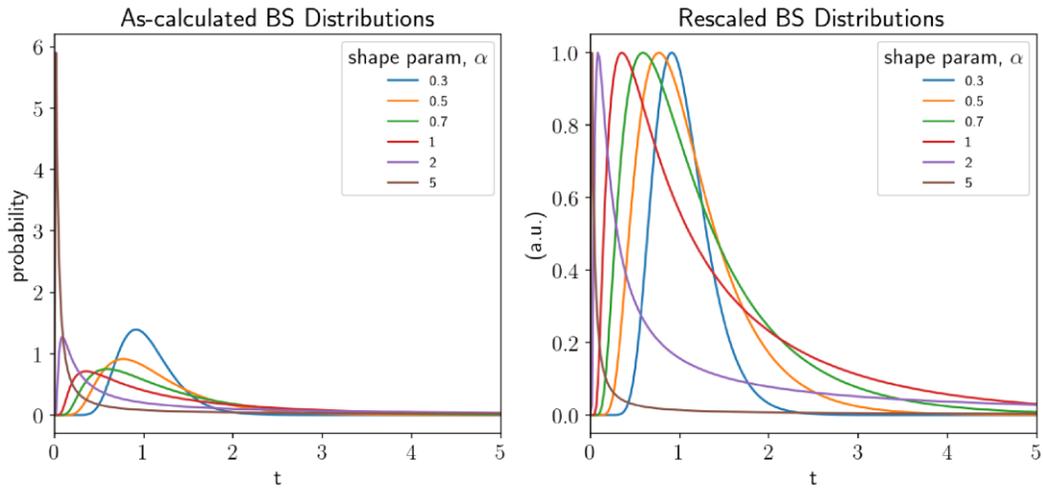

## Scale parameter

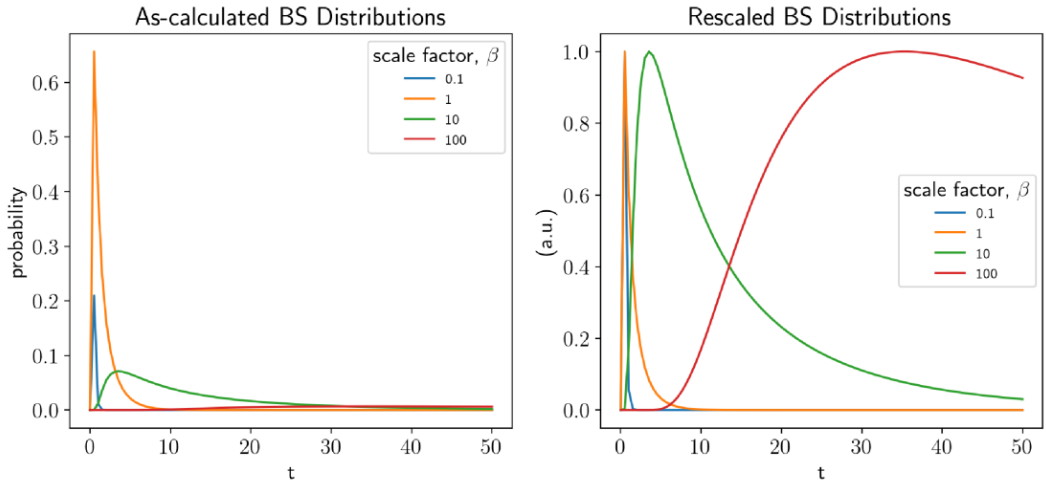

**Figure S10** Illustration of the effect of the shape and scale parameters on the Birnbaum-Saunders distribution.

**References**
(1) Balakrishnan, N.; Kundu, D. Birnbaum-Saunders Distribution: A Review of Models, Analysis, and Applications. *Applied Stochastic Models in Business and Industry* **2019**, *35* (1), 4–49. https://doi.org/10.1002/asmb.2348.